\documentclass[11pt,english]{article}

\usepackage[dvipdfm]{graphicx}
\usepackage{amsmath}
\usepackage{amssymb}
\usepackage{bm}

\newcommand {\La} {{\cal L}}

\newcommand {\kfermi} {k_{\scriptscriptstyle F}}

\newcommand {\Mstar} {M^{\ast}}

\newcommand {\rhoB} {\rho_{\scriptscriptstyle B}}

\def\slashchar#1{\setbox0=\hbox{$#1$} 
\dimen0=\wd0 \setbox1=\hbox{/} \dimen1=\wd1
\ifdim\dimen0>\dimen1 \rlap{\hbox to \dimen0{\hfil/\hfil}} #1
\else  \rlap{\hbox to \dimen1{\hfil$#1$\hfil}} / \fi}

\makeatletter

\@addtoreset{equation}{section}
\makeatother

\begin{document}

\pagestyle{empty}

\baselineskip=17.0pt

\centerline{Chiral-particle Approach to Hadrons}

\centerline{in an Extended Chiral ($\sigma,\pi,\omega$) Mean-Field Model}

\vspace{1.0cm}

\centerline{Schun~T.~Uechi\footnote{E-mail: suechi@rcnp.osaka-u.ac.jp} and Hiroshi Uechi\footnote{E-mail: uechi@ogu.ac.jp}}
\vspace{0.5cm}

\centerline{$^1$Research Center for Nuclear Physics ($RCNP$), Osaka University, Ibaraki, Osaka 567-0047 \quad Japan}

\centerline{$^2$Department of Distributions and Communication Sciences, Osaka Gakuin University}
\centerline{2-36-1  Kishibe-minami,  Suita,  Osaka  564-8511 \quad Japan}

\vspace{1.5cm}

\baselineskip=13.0pt

\centerline{\large{Abstract}}

The chiral nonlinear ($\sigma,\pi,\omega$) mean-field model is an
extension of the conserving nonlinear (nonchiral) $\sigma$-$\omega$
hadronic mean-field model which is thermodynamically consistent,
relativistic and Lorentz-covariant mean-field theory of hadrons.  In the
extended chiral ($\sigma,\pi,\omega$) mean-field model, all the masses
of hadrons are produced by chiral symmetry breaking mechanism, which is
different from other conventional chiral partner models.  By comparing
both nonchiral and chiral mean-field approximations, the effects of
chiral symmetry breaking to the mass of $\sigma$-meson, coefficients of
nonlinear interactions, coupling ratios of hyperons to nucleons and
Fermi-liquid properties are investigated in nuclear matter, hyperonic
matter, and neutron stars.



\pagestyle{plain}
\setcounter{page}{1}

\section{Introduction}

A renormalizable quantum field theory based on hadronic degrees of
freedom provides us with an intuitively and physically accessible
approach from finite nuclei to infinite nuclear matter, {\it e.g.}, high
density nuclear and hyperonic matter such as neutron (hadronic)
stars~$\cite{WAL}-\cite{SCH2}$.  The linear neutral scalar and vector
($\sigma$,$\omega$), nonlinear ($\sigma$,$\omega$), nonlinear
($\sigma$,$\omega$,$\rho$) mean-field models are actively studied and
applied to finite and infinite hadronic many-body systems.  Though
hadronic mean-field models render nuclear and astronomical phenomena
readily understandable, they are strongly interacting particles, which
makes hadronic approach much complicated.  One may investigate the
hadronic system by starting from quantum chromodynamics (QCD), but
because of strong interactions, it is complicated in the nuclear energy
domain so that one is led to introduce certain effective hadronic models
to simulate strong interactions of hadrons~$\cite{SER}$.

The hadronic mean-field models must be constructed to reproduce binding
energy at saturation of symmetric nuclear matter (assumed as $-15.75$
MeV at $\rho_0 = 0.148$ fm$^{-3}$ or $\kfermi = 1.30$ fm$^{-1}$ in the
current calculation), which is one of fundamental requirements for
nuclear physics.  The pressure must vanish at saturation ($p=0$), and
simultaneously, the self-consistent single particle energy,$E(\kfermi)$,
must be obtained by the functional derivative of energy density with
respect to baryon density, $\delta {\cal E}/\delta \rhoB = E(\kfermi)$,
as a dynamical constraint for any employed approximation.  The energy
density and pressure must maintain a thermodynamic relation, such as
${\cal E} + p = \mu \rhoB$ (at $T=0$), to be a self-consistent
approximation for nuclear matter.  In terms of dynamical quantities, the
self-consistent requirement can be stated that Green function,
self-energy and energy density must maintain conditions of conserving
approximations, termed as {\it thermodynamic consistency}.
Thermodynamic consistency is explicitly expressed as the requirement
that functional derivatives of energy density with respect to
self-energies must vanish, $\delta {\cal E}/\delta \Sigma = 0
$~$\cite{UEC1}$, which becomes equivalent to Landau's hypothesis of
quasiparticles and the fundamental requirement of density functional
theory~$\cite{WKS}-\cite{UEC3}$.

Although the linear and nonlinear ($\sigma$,$\omega$,$\rho$) mean-field
models appropriately simulate properties of symmetric nuclear matter and
neutron stars, they have many free parameters, masses and nonlinear
coupling constants, coming from meson fields and nonlinear interactions.
The upper bounds of values of nonlinear coefficients are confined by
maintaining conditions of thermodynamic consistency to an employed
approximation~$\cite{UEC2}$ and reproducing empirical data.  Nonlinear
coefficients are bounded for thermodynamically consistent
approximations, and this is discussed as a manifestation of {\it
naturalness} for self-consistent approximations.  However, it is beyond
the linear and nonlinear mean-field models to answer the reason why
coupling constants for nonlinear interactions are needed and restricted
with such strengths.  The chiral mean-field model reveals important
meanings and strength of nonlinear interactions~$\cite{W95}-\cite{S07}$.
The current chiral ($\sigma,\pi,\omega$) mean-field approximation
provides with the followings:

1. Generations of hadron masses by way of chiral symmetry breaking
correspondingly produce coefficients of nonlinear meson interactions.
It indicates that the fundamental requirement at nuclear matter
saturation is directly related with experimental values of hadron masses
($M_N, m_{\pi}, m_{\omega}, m_{\rho}, \cdots$).  In mean-field (Hartree)
approximation, pion contributions vanish, and $\sigma$-meson compensates
for attractive contributions expected to be given by pions at saturation
density. Hence, the saturation property determines the mass of sigma
meson, $m_{\sigma}$.

2. The coupling constants for hyperons are important to study phase
transitions from $\beta$-equilibrium ($n,p,e$) asymmetric nuclear matter
to ($n,p,H_1,e$) hyperonic matter, binding energy of pure-hyperon matter
and masses of hadronic stars.  It is found that $\Lambda$-hyperon
coupling ratio to nucleon, $r^{\omega}_{\Lambda N} = g_{\omega
\Lambda}/g_{\omega N}$, is expected to be $r^{\omega}_{\Lambda N} \sim
1.0$ by the requirement of thermodynamic consistency~$\cite{SCH,UEC3}$,
whereas the SU(6) quark model for hadrons demands
$r^{\omega}_{\Lambda N} \sim 2/3$, or $1/3$~$\cite{JSI,GSY}$.  The differences of
$r^{\omega}_{\Lambda N}$ result in significant discrepancies in
effective masses of hadrons, onset densities of nucleon-hyperon phase
transitions, saturation properties of hyperons, and masses of hadron
stars~$\cite{SCH}$.  If the current chiral ($\sigma, \pi, \omega$)
mean-field model is applied to phase transition to $\beta$-equilibrium
Lambda matter ($n,p,\Lambda,e$), it deduces that $r^{\sigma}_{\Lambda N}
= g_{\sigma \Lambda}/g_{\sigma N} = M_{\Lambda}/M_N \approx 1.187$ (see,
sec.~3), which is consistent with the analysis of the conserving, nonchiral
($\sigma,\omega,\rho$) mean-field approximation.

The current extended chiral ($\sigma,\pi,\omega$) mean-field model
starts from a lagrangian without hadron masses and generates
all the hadron masses by way of chiral symmetry breaking.  This is
different from other chiral mean-field models which introduce the
isoscalar-vector particle, $\omega$, externally in order to produce
repulsive interaction and saturation mechanism.  The current chiral
($\sigma,\pi,\omega$) mean-field model produces masses of $\sigma$,
$\pi$ and $\omega$ particles by chiral symmetry breaking mechanism.  The
chiral symmetric interaction and chiral breaking, binding energy are
discussed in sec.~2 and Fermi-liquid properties of nuclear matter, such
as incompressibility and symmetry energy, $K$ and $a_4$, and numerical
results are shown in sec.~3.

The vacuum fluctuation corrections to the chiral ($\sigma,\pi,\omega$)
mean-field approximation, applications to $\beta$-equilibrium ($n,p,e$)
asymmetric nuclear matter and properties of hadron (neutron) stars are
discussed in sec.~4.  The phase transition from symmetric nuclear
matter to $\beta$-equilibrium hyperon matter, ($n,p,H_1,e$), and
important results on coupling ratios given by chiral symmetry breaking
are also discussed, and concluding remarks are in sec.~5.

\section{An extended chiral $\sigma$,$\pi$,$\omega$ nonlinear mean-field approximation}

The conventional chiral mean-field models for hadrons suppose that the
lagrangian with interaction potential, $V(\sigma^{2}+\bm{\pi}^{2})$,
should be invariant under the chiral transformation and constrain only
$\sigma$ and $\bm{\pi}$ mesons as a {\it chiral partner}.  Moreover, a
massive isoscalar vector field $\omega_{\mu}$ is input externally to
supply repulsive nuclear-nuclear interactions as in
QHD-I${\cite{WAL,Serot_1997xg}}$.  The conventional chiral mean-field
models for hadrons exhibit that when chiral symmetry breaking parameter
vanishes, the masses $m_{\sigma}$ and $m_{\pi}$ vanish:
$m_{\sigma}\rightarrow 0$, $m_{\pi}\rightarrow 0$, whereas
$m_{\omega}\rightarrow \hspace*{-0.4cm}/ \hspace*{0.3cm}0$.

We introduce an extended chiral symmetric mean-field lagrangian for
hadrons with the interaction potential,
$V(\sigma^{2}+\bm{\pi}^{2}-a\omega_{\mu}^{2})$.  The lagrangian is
invariant under the chiral transformation and produces all hadron masses
and nonlinear mean-field interactions by way of chiral symmetry
breaking. The parameter $a$ is constant, which will be identified as
$m_{\omega}^{2}/m_{\pi}^{2} \sim 31.65$ at nuclear domain, after the
chiral symmetry breaking.  Therefore, the current extended chiral
mean-field model generates $\omega$-meson as {\it chiral particles} such
that all the meson masses are required to vanish simultaneously:
$m_{\sigma}\rightarrow 0$, $m_{\pi}\rightarrow 0$, and
$m_{\omega}\rightarrow 0$ when the chiral breaking parameter vanishes,
$\varepsilon\rightarrow 0$.  In other words, we assume that all the
hadron masses ($M_N, m_{\sigma}, m_{\pi}, m_{\omega}$) and nonlinear
interactions be generated by the lagrangian with interaction potential
$V(\sigma^{2}+\bm{\pi}^{2}-a\omega_{\mu}^{2})$ under the chiral symmetry
breaking mechanism.

The current extended chiral mean-filed model that produces all the hadron masses
$(M_N,m_{\sigma},m_{\pi},m_{\omega})$ with the chiral symmetry breaking is based on a
relativistic chiral ($\sigma,\bm{\pi},\omega$) model discussed by Walecka, Serot
and others${\cite{W95}-\cite{S07}}$.  The extended chiral $(\sigma,\bm{\pi},\omega)$ lagrangian is
\begin{equation}
\begin{split}
\La=&\bar\psi\left[\gamma_{\mu}(i\partial^{\mu}-g_{\omega}\omega^{\mu})+g(\sigma+i\gamma_{5}{\bm\tau}\cdot{\bm\pi})\right]\psi\\
&+\frac{1}{2}(\partial_{\mu}\sigma\partial^{\mu}\sigma+\partial_{\mu}\bm{\pi}\cdot\partial^{\mu}\bm{\pi})-\frac{1}{4}F_{\mu\nu}F^{\mu\nu}-V(\sigma^{2}
+\bm{\pi}^{2}-a\omega_{\mu}^{2})-\delta\La_{csb}, \label{eqn:lag}
\end{split}
\end{equation}
where $\delta\La_{csb}=\varepsilon\sigma$ is the chiral symmetry
breaking term.  The nucleon is
$\psi=\begin{pmatrix}\psi_{p}\\\psi_{n}\end{pmatrix}$, and
$\sigma,\bm{\pi},\omega_{\mu}$ are neutral scalar meson, isovector pion
and neutral isovector omega meson fields, respectively,
and $F_{\mu\nu}=\partial_{\mu}\omega_{\nu}-\partial_{\nu}\omega_{\mu}$ is
for the vector-isoscalar $\omega$-meson.  Note that there are no baryon
and meson masses in the lagrangian $\eqref{eqn:lag}$, and baryons and
mesons are coupled as $g_{\omega}
\bar{\psi}\gamma_{\mu}\omega^{\mu}\psi$, and
$g\bar{\psi}(\sigma+i\gamma_{5}{\bm\tau}\cdot{\bm\pi})\psi$.  The
coupling constant, $g$, is pion-nucleon (and $\sigma$-nucleon) coupling
constant to be required from invariance under the chiral transformation
($g_{\sigma}=g_{\pi}=g$ is assumed).
We introduce the chiral-invariant potential of the form:
\begin{equation}
 V(\sigma^{2}+\bm{\pi}^{2}-a\omega_{\mu}^{2})=\frac{\lambda}{4}\left[\sigma^{2}+\bm{\pi}^{2}-a\omega_{\mu}^{2}\right]^{2},\label{eqn:pv}
\end{equation}
where $\lambda\neq 0$ and $a>0$ are constants determined in the ground
state after the chiral-symmetry breaking.  Hence, the free parameters of
the current chiral mean-field model are $g$, $g_{\omega}$ and $\lambda$.
Note that $(\sigma,\bm{\pi},\omega)$ mesons make the lagrangian chiral
invariant all together, not by way of the chiral-invariance generated by
$(\sigma,\bm{\pi})$-chiral partner, and in this sense, we call
$(\sigma,\bm{\pi},\omega)$ mesons as chiral particles.

The current chiral lagrangian is invariant under the following gauge
transformations
\begin{equation}
\begin{split}
&\delta\psi=\frac{i}{2}\bm\epsilon\cdot\bm\tau\gamma_{5}\psi,\\
&\delta\bm{\pi}=-\bm{\epsilon}\sigma,\\
&\delta\sigma=\bm{\epsilon}\cdot\bm{\pi},
\end{split}
\end{equation}
and $\bm{\epsilon}$ is supposed to be an infinitesimal value, and the
$\omega$ meson is invariant under the gauge transformation:
$\delta\omega_{\mu}=0$.  After the chiral symmetry breaking, the
interaction potential is given in the new ground state as,
\begin{equation}
\begin{split}
 V&=\frac{\lambda}{4}\left[\left(\sigma^{2}+\bm{\pi}^{2}-a\omega_{\mu}^{2}\right)-v^{2}\right]^{2}+\delta\La_{csb}\\
&=\frac{\lambda}{4}\left[\left(\sigma^{2}+\bm{\pi}^{2}-a\omega_{\mu}^{2}\right)-v^{2}\right]^{2}+\varepsilon\sigma, \label{eqn:VCSB}
\end{split}
\end{equation}
where $\lambda$, $v$, $a$ and $\varepsilon$ are constants determined at
the ground state.

The mesons are excited from the new ground state as follows,
\begin{equation}
\begin{split}
 \sigma&\rightarrow\left<\sigma\right>+\phi, \\
 \bm{\pi}&\rightarrow\left<\bm{\pi}\right>+\bm{\pi}, \\
 \omega_{\mu}&\rightarrow\left<\omega_{\mu}\right>+\omega_{\mu}, \label{eqn:MES}
\end{split}
\end{equation}
where $\left<\sigma\right>$, $\left<\bm{\pi}\right>$ and
$\left<\omega_{\mu}\right>$ are values for the meson fields in the
vacuum defined by minimization of $\eqref{eqn:VCSB}$ with respect to
$\sigma,\bm{\pi}$, and $\omega_{\mu}$.  The interaction potential $V$
has the following form at the ground state in the new vacuum,
\begin{equation}
V=\frac{\lambda}{4}\left[\left(\left<\sigma\right>^{2}+\left<\bm{\pi}\right>^{2}-a\left<\omega_{\mu}\right>^{2}\right)-v^{2}\right]^{2}+\varepsilon\left<\sigma\right>,\label{eqn:min}
\end{equation}
and the minimization conditions give
\begin{equation}
\begin{split}
 \frac{\partial V}{\partial\left<\sigma\right>}&=\lambda\left<\sigma\right>\left[(\left<\sigma\right>^{2}+\left<\bm{\pi}\right>^{2}-a\left<\omega_{\mu}\right>^{2})-v^{2}\right]+\varepsilon=0, \\
 \frac{\partial V}{\partial\left<\bm{\pi}\right>}&=\lambda\left<\bm{\pi}\right>\left[(\left<\sigma\right>^{2}+\left<\bm{\pi}\right>^{2}-a\left<\omega_{\mu}\right>^{2})-v^{2}\right]=0, \\
 \frac{\partial V}{\partial\left<\omega_{\mu}\right>}&=\lambda a\left<\omega_{\mu}\right>\left[(\left<\sigma\right>^{2}+\left<\bm{\pi}\right>^{2}-a\left<\omega_{\mu}\right>^{2})-v^{2}\right]=0 \ .
 \label{eqn:mvac}
\end{split}
\end{equation}
The conditions, $\lambda\neq 0$ and $\varepsilon\neq 0$, lead to $\left<\sigma\right> \equiv\sigma_{0} \neq 0$,
$[\left<\sigma\right>^{2}+\left<\bm{\pi}\right>^{2}-a\left<\omega_{\mu}\right>^{2}-v^{2}]\neq 0$, and
\begin{equation}
 \left<\bm{\pi}\right>=0,\hspace{0.5cm}\left<\omega_{\mu}\right>=0 \ . \label{eqn:fvac}
\end{equation}
The ground state value, $\sigma_{0}$, is then defined as,
\begin{equation}
\left<\sigma\right>\equiv\sigma_{0}=-\frac{M}{g} .
\end{equation}
By expanding the interaction potential
$V(\sigma^{2}+\bm{\pi}^{2}-a\omega_{\mu}^{2})$, the terms in
$\eqref{eqn:min}$ are collected as follows:

(1) Constant terms are
\begin{equation}
 V_{0}=\frac{\lambda}{4}(\sigma_{0}^{2}-v^{2})^{2}+\varepsilon\sigma_{0}.
\end{equation}

(2) The terms linear in $\phi$ are
\begin{equation}
 V_{1}=\left\{\lambda\sigma_{0}(\sigma_{0}^{2}-v^{2})+\varepsilon \right\}\phi = 0 \ .
\end{equation}
This expression vanishes because of the minimization conditions,
$\eqref{eqn:mvac}$ and $\eqref{eqn:fvac}$.

(3) The terms quadratic in $\bm{\pi}$ are
\begin{equation}
V_{2}=-\frac{1}{2}\frac{\varepsilon}{\sigma_{0}}{\bm{\pi}^{2}}=\frac{1}{2}\frac{g\varepsilon}{M}{\bm{\pi}^{2}}\equiv\frac{1}{2}\mu_{2}^{2}{\bm{\pi}}^{2}.
\end{equation}

(4) The terms quadratic in $\omega_{\mu}$ are derived in the same way as,
\begin{equation}
V_{3}=-\frac{1}{2}\frac{g\varepsilon}{M}a\omega_{\mu}^{2}\equiv -\frac{1}{2}\mu_{3}^{2}\omega_{\mu}^{2}.
\end{equation}

(5) The terms quadratic in $\phi$ are
\begin{equation}
V_{4}=\frac{\lambda}{2}(\sigma_{0}^{2}-v^{2})\phi^{2}+\lambda\sigma_{0}^{2}\phi^{2}\equiv\frac{1}{2}\mu_{1}^{2}\phi^{2}.
\end{equation}
and $\lambda$ is given by
\begin{equation}
\lambda\equiv\frac{1}{2}\left(\frac{g}{M}\right)^{2}(\mu_{1}^{2}-\mu_{2}^{2}).
\end{equation}
\begin{figure}[htpb]
\begin{minipage}[t]{0.48\textwidth}
\begin{center}
\includegraphics[height=.25\textheight]{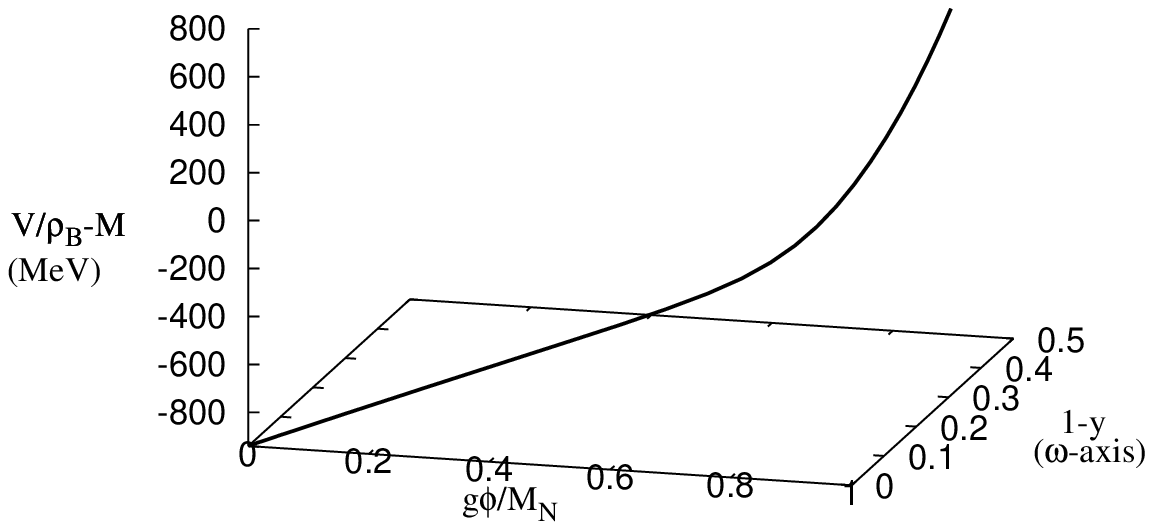}
\label{fig1a}
\end{center}\vspace{-0.5cm}
Fig.1.  The interaction potential $V$ defined by meson sectors.  The field $\phi$ produces attractive interaction at low
densities.  The $\omega$-axis is written by the variable $1-y$, where $y=(m_{\omega}^2/g_{\omega}\rhoB) \omega_0$, and
$\omega$-field produces repulsive interaction at high densities.
\end{minipage}
\hspace{0.5cm}
\begin{minipage}[t]{0.48\textwidth}
\begin{center}
\includegraphics[height=.25\textheight]{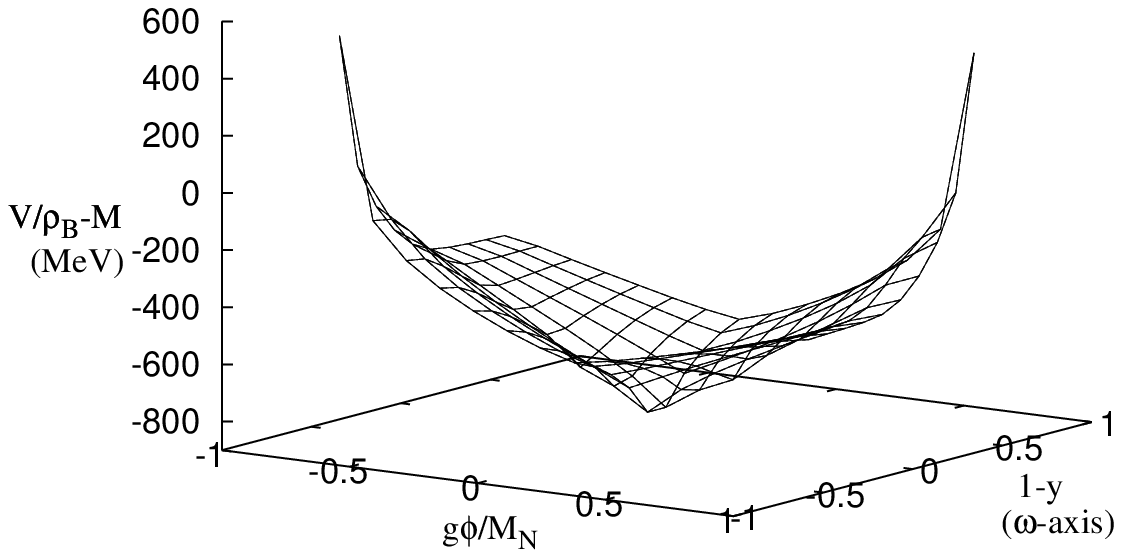}
\label{fig2a}
\end{center}\vspace{-0.5cm}
Fig.2.  The 3-dimensional image of interaction potential $V$.  The origin is set in the center of $\phi$-$\omega$ plain.
\end{minipage}
\end{figure}

(6) The remaining cubic and quartic interactions of the meson fields
$(\phi,{\bm{\pi}},\omega)$ are then given by
\begin{equation}
 \begin{split}
  V_{5}+V_{6}+V_{7}&=\frac{\lambda}{4}\left[4\sigma_{0}\phi(\phi^{2}+{\bm{\pi}}^{2}-a\omega_{\mu}^{2})+(\phi^{2}+{\bm{\pi}}^{2}-a\omega_{\mu}^{2})^{2}\right]\\
&=\frac{1}{2}(\mu_{1}^{2}-\mu_{2}^{2})\left[\left(\frac{g}{2M}\right)^{2}(\phi^{2}+{\bm{\pi}}^{2}-a\omega_{\mu}^{2})^{2}-2\left(\frac{g}{2M}\right)\phi(\phi^{2}+{\bm{\pi}}^{2}
-a\omega_{\mu}^{2})\right].
\end{split}
\end{equation}
A collection of these terms then yields the interaction potential $V$
written as,
\begin{equation}
\begin{split}
 V=&\frac{1}{2}\mu_{1}^{2}\phi^{2}+\frac{1}{2}\mu_{2}^{2}{\bm{\pi}}^{2}-\frac{1}{2}\mu_{3}^{2}\omega_{\mu}^{2}\\
&+\frac{1}{2}(\mu_{1}^{2}-\mu_{2}^{2})\left[\left(\frac{g}{2M}\right)^{2}(\phi^{2}+{\bm{\pi}}^{2}-a\omega_{\mu}^2)^2
-2\left(\frac{g}{2M}\right)\phi(\phi^2+{\bm{\pi}}^{2}-a\omega_{\mu}^2)\right].
\end{split}
\end{equation}
The lagrangian density~$\eqref{eqn:lag}$ with the generation of hadron
masses by spontaneous symmetry breaking finally takes the following
form:
\begin{equation}
\begin{split}
\La_{csb}=&\bar\psi\left[\gamma_{\mu}(i\partial^{\mu}-g_{\omega}\omega^{\mu})-\left\{M-g(\phi+i\gamma_{5}{\bm\tau}\cdot{\bm\pi})\right\}\right]\psi\\
&+\frac{1}{2}(\partial_{\mu}\phi\partial^{\mu}\phi-\mu_{1}^{2}\phi^{2})+\frac{1}{2}(\partial_{\mu}\bm{\pi}\cdot\partial^{\mu}\bm{\pi}-\mu_{2}^{2}\bm{\pi}^{2})-\frac{1}{4}F_{\mu\nu}F^{\mu\nu}+\frac{1}{2}\mu_{3}^{2}\omega_{\mu}^{2}\\
&-\frac{1}{2}(\mu_{1}^{2}-\mu_{2}^{2})\left[\left(\frac{g}{2M}\right)^{2}(\phi^{2}+{\bm{\pi}}^{2}-a\omega_{\mu}^{2})^{2}-2\left(\frac{g}{2M}\right)\phi(\phi^{2}+{\bm{\pi}}^{2}-a\omega_{\mu}^{2})\right]+{\rm constant} \ .
\end{split}
\end{equation}
The parameters are identified to be: $\mu_{1}=m_{\sigma},
\mu_{2}=m_{\pi}, \mu_{3}=m_{\omega}$, and $a \equiv
m_{\omega}^{2}/m_{\pi}^{2} \sim 31.65$ in nuclear domain.

The chiral $(\sigma,{\bm{\pi}},\omega)$ mean-field approximation is
defined by replacing meson quantum fields with classical fields:
$\hat{\phi}\rightarrow\phi_{0}$,
$\hat{\omega}_{\mu}=(\hat{\omega}_{0},\bm
\omega)\rightarrow(\omega_{0},\bm 0)$, they are constants independent of
$x_{\mu}$.  The spacial part of the vector field
$\left<{\bm{\omega}}\right>$ should vanish by the requirement of
rotational invariance of static and homogeneous nuclear
matter~$\cite{WAL}$, and in addition, $\pi$-meson contributions vanish
in the (mean-field) Hartree approximation.  The chiral mean-field
lagrangian is given by
\begin{equation}
\begin{split}
\La_{csb}=&\bar\psi\left[\gamma_{\mu}(i\partial^{\mu}-g_{\omega}\omega_{0})-(M-g\phi_{0}) \right]\psi\\
&-\frac{1}{2}m_{\sigma}^{2}\phi_{0}^{2}+\frac{1}{2}m_{\omega}^{2}\omega_{0}^{2}-\frac{1}{2}(m_{\sigma}^{2}-m_{\pi}^{2})\left[\left(\frac{g}{2M}\right)^{2}(\phi_{0}^{2}-a\omega_{0}^{2})^{2}-2\left(\frac{g}{2M}\right)\phi_{0}(\phi_{0}^{2}-a\omega_{0}^{2})\right].
\end{split}
\end{equation}
The equations of motion for the scalar and vector mesons are given by
\begin{equation}
m_{\sigma}^2 \phi_0^2- \frac{g}{2M}(m_{\sigma}^2-m_{\pi}^2) \left(3\phi_0^2+2a\phi_0 \omega_0^2-2\frac{g}{2M}\phi_0^3 \right)
= g \rho_s^{\ast}, \label{eqn:ES}
\end{equation}
\begin{equation}
m_{\omega}^2 \omega_0-2\frac{g}{2M}(m_{\sigma}^2-m_{\pi}^2)\left(a\phi_0 \omega_0+\frac{g}{2M}a^{2}\omega_{0}^{3}-\frac{g}{2M}a\phi_0^2\omega_{0}\right)
= g_{\omega}\rho_{B},
\end{equation}
where $\rho_{\sigma}^{\ast}$ is the scalar source, and $\rhoB$ is the
baryon density: $\rho_{B} = \displaystyle \sum_B k_{F_B}^3/3\pi^2$,
where $k_{F_B}$ is a baryon Fermi-momentum.  The energy density and
pressure can be derived from energy momentum
tensor~$\cite{Serot_1997xg,UEC2}$:
\begin{equation}
\begin{split}
{\cal E}=&\sum_{B=n,p}\frac{2}{(2\pi)^{3}}\int^{k_{F_{B}}} d^{3}k E_{B}(k)+\frac{1}{2}m_{\sigma}^2\phi_0^2-\frac{g}{2M}(m_{\sigma}^2-m_{\pi}^2)\left(\phi_0^3
-\frac{1}{2}\frac{g}{2M}\phi_0^4 \right)\\
&-\frac{1}{2}m_{\omega}^2\omega_0^2+\frac{g}{2M}(m_{\sigma}^{2}-m_{\pi}^{2})a\left(\phi_0\omega_0^2+\frac{1}{2}\frac{g}{2M} a\omega_0^4
-\frac{g}{2M}\phi_0^2 \omega_0^2 \right),
\end{split}
\end{equation}
\begin{equation}
\begin{split}
p=&\sum_{B=n,p}\frac{1}{3}\frac{2}{(2\pi)^3}\int^{k_{F_{B}}} d^{3}k \frac{k^{2}}{E_{B}^{\ast}(k)}-\frac{1}{2}m_{\sigma}^{2}\phi_{0}^{2}+\frac{g}{2M}(m_{\sigma}^2
-m_{\pi}^{2})\left(\phi_{0}^{3}-\frac{1}{2}\frac{g}{2M}\phi_0^4 \right)\\
&+\frac{1}{2}m_{\omega}^2 \omega_{0}^{2}-\frac{g}{2M}(m_{\sigma}^2-m_{\pi}^2) a\left(\phi_{0}\omega_0^2+\frac{1}{2}\frac{g}{2M}a\omega_{0}^{4}
-\frac{g}{2M}\phi_0^2 \omega_0^2 \right),
\end{split}
\end{equation}
where
$E_{B}(k)=E_{B}^{\ast}(k)+\Sigma_{\omega}^{0}=\sqrt{k^{2}+M_{B}^{\ast
2}}-g_{\omega}\omega_{0}$.  The scalar source $\rho_{\sigma}^{\ast}$ is
derived from the functional derivative with respect to
$\phi_0$~$\cite{UEC2}$:
\begin{equation}
\rho_s^{\ast} = \sum_B \frac{1}{\pi^2 }\int^{k_{F_{B}}}\! q^2 d q \frac{\Mstar}{E^{\ast}(q)}
-\frac{1}{2M}(m_{\sigma}^{2}-m_{\pi}^2) a \omega_0^2 \ . \label{eqn:SS}
\end{equation}
The self-consistent effective masses of hadrons are determined by
satisfying conditions of thermodynamic consistency~$\cite{UEC2}$:
\begin{equation}
\begin{split}
 M_N^{\ast} &= M - g \phi_0, \\
 m_{\sigma}^{\ast 2} &= m_{\sigma}^2 - \frac{3g}{2M}(m_{\sigma}^2-m_{\pi}^2)\phi_0 + 2(m_{\sigma}^2-m_{\pi}^2)(\frac{g}{2M})^2
\left( \phi_0^2 - a \omega_0^2 \right), \\
 m_{\omega}^{\ast 2} &= m_{\omega}^2 - \frac{g}{M}a(m_{\sigma}^2-m_{\pi}^2)\phi_0 + 2a(m_{\sigma}^2-m_{\pi}^2)(\frac{g}{2M})^2
 -2a^2(m_{\sigma}^2-m_{\pi}^2)(\frac{g}{2M})^2 \omega_0^2 \ , \label{eqn:CHM}
\end{split}
\end{equation}
and self-consistent scalar and vector self-energies are given
by~$\cite{UEC2}$:
\begin{equation}
\Sigma^s = -\frac{g^2}{m_{\sigma}^{\ast 2}} \rho_s^{\ast} \ , \qquad
\Sigma_{\omega}^{\mu} = - \frac{g_{\omega}^2}{m_{\omega}^{\ast 2}} \rhoB \delta_{\mu, 0} \ , \label{eqn:S0}
\end{equation}
where $m_{\sigma}^{\ast}$ and $m_{\omega}^{\ast}$ are effective masses
of $\sigma$ and $\omega$ mesons.

The interaction potential defined by meson sectors is shown in Fig.~1
and a three-dimensional image of the interaction potential is shown in
Fig.~2.  In the current chiral mean-field approximation, the interaction
potential is self-consistently constructed by $\sigma$ and $\omega$
mesons; $\sigma$-meson produces attractive interaction at low densities,
whereas $\omega$-meson mainly generates repulsive contributions at high
densities.  The energy density and pressure satisfy, ${\cal E} + p = \mu
\rhoB$ and $\mu = E(\kfermi)$ in all densities.  The binding energies of
symmetric nuclear matter and ($n,p,e$) asymmetric matter are shown in
Fig.~3.
\begin{figure}[t]
\begin{center}
\includegraphics[height=.25\textheight]{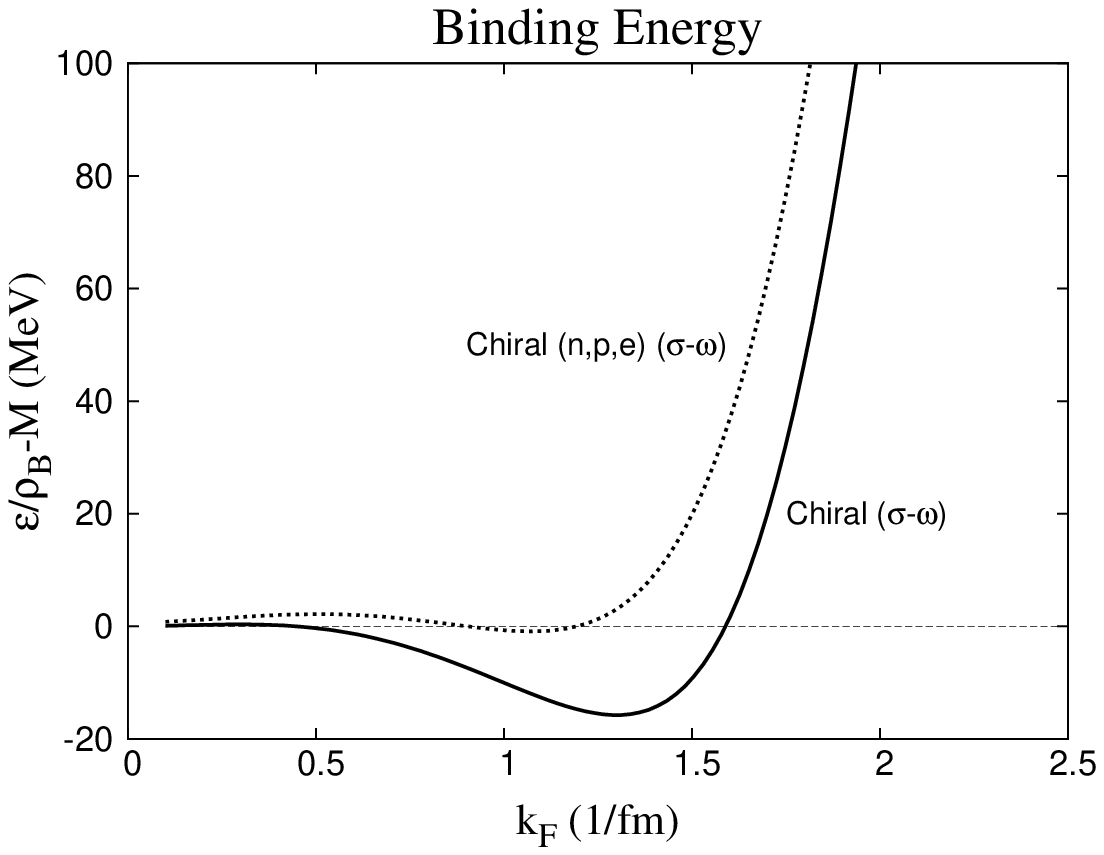}
\label{fig3}
\end{center}\vspace{-0.5cm}
Fig.3.  The binding energies of isospin symmetric ($n,p$) and isospin
 asymmetric ($n,p,e$) matter.  Note that ${\cal E}/\rhoB = E(\kfermi)$
is exactly satisfied at saturation density, $\rhoB = 0.148$ fm$^{-3}$.
\end{figure}
\begin{figure}[htpb]
\begin{center}
\includegraphics[height=.25\textheight]{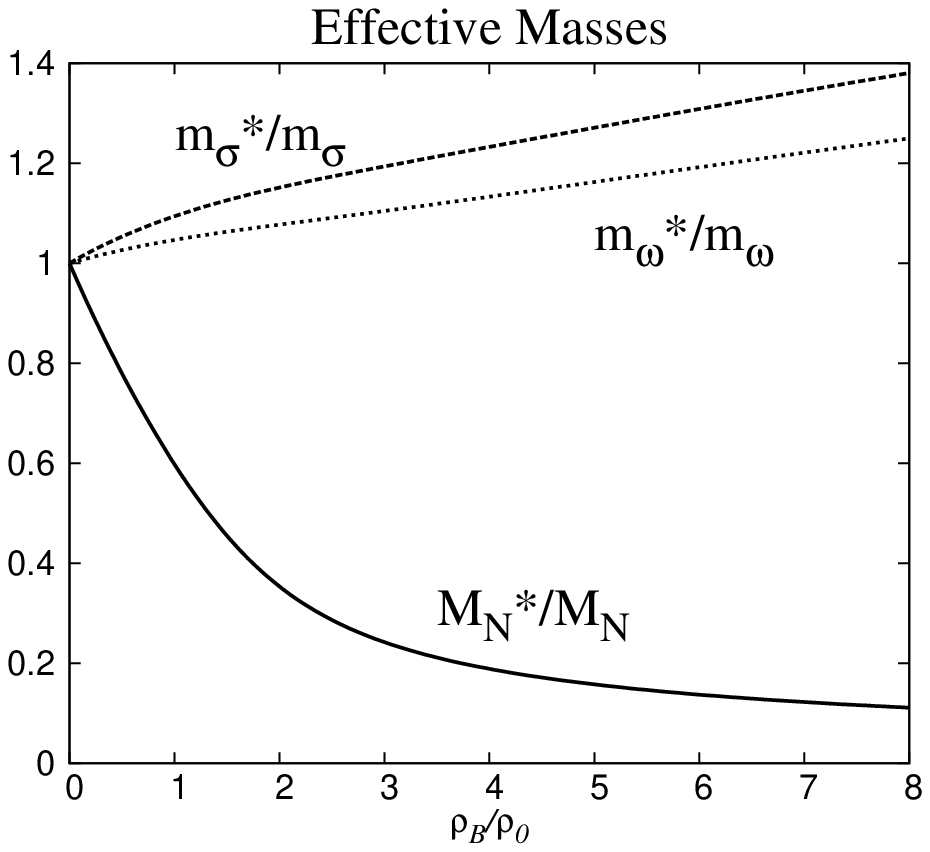}
\label{fig4}
\end{center}\vspace{-0.5cm}
Fig.4.  Effective masses of nucleon, $M_N^{\ast}/M_N$, and mesons,
 $m^{\ast}_{\sigma}/m_{\sigma}$ and $m^{\ast}_{\omega}/m_{\omega}$.  The
 qualitative behavior of effective masses are consistent with those
 derived from nonlinear, nonchiral ($\sigma,\omega,\rho$) mean-field
approximation.
\end{figure}

\section{Fermi liquid properties at nuclear matter saturation}

The chiral ($\sigma,\pi,\omega$) mean-field model exhibits remarkable
properties when it is compared to the nonchiral, nonlinear
($\sigma,\omega$,$\rho$) mean-field model.  The nonchiral mean-field
model is applied to ($n,p$) symmetric, ($n,p,e$) asymmetric, ($n,p,H,e$)
hyperonic matter, and neutron stars~$\cite{SCH}$.  Although the
nonchiral model reasonably simulates properties of nuclear and neutron
matter, it has many parameters such as masses and coupling constants.
The upper and lower bound values of coupling constants and effective
masses of hadrons are constrained by empirical data and self-consistent
conditions to approximations.  The nonlinear nonchiral mean-field
approximations could not clearly explain the reason why values of
nonlinear coupling constants are bound in a characteristic
way~$\cite{SCH}$.  The chiral ($\sigma,\pi,\omega$) mean-field model
clarifies relations among nonlinear coupling constants, hadron masses
and observables.

\begin{table}[b]
\noindent{Table~1.\quad Coupling constants and Fermi-liquid properties of nuclear matter}
\begin{center}
\arrayrulewidth=0.5pt
\doublerulesep=0pt
\begin{tabular}{ccccc} \cline{1-3}
$g$ & $g_{\omega}$  & $m_{\sigma}$  \\
 2.4095 & 13.4232 & 120.0 &        \\ \cline{1-3}
\hline \hline
$\Mstar_N/M_N$ & $m_{\sigma}^{\ast}/m_{\sigma}$ & $m_{\omega}^{\ast}/m_{\omega}$ & $K$ (MeV) & $a_4$ (MeV) \\
 0.60 & 1.09 & 1.04 & 371 & 17.4 \\
\hline
$M_{max}$ & ${\cal E}_{c}$ & $I$ & $R$ (km) \\
 2.60 & 1.58 & 418 & 12.8 \\
\cline{1-4}
\end{tabular}
\end{center}
In the current chiral mean-field approximation, the masses of $\pi$ and
$\omega$ mesons are identified as, $\mu_2 = m_{\pi} = 139.0$ MeV and
$\mu_3 = m_{\omega} = 783.0$ MeV, in the nuclear domain.  Hence,
adjustable parameters are only $g$, $g_{\omega}$, and $m_{\sigma}$.  The
effective masses, $K$ and $a_4$, are values at saturation of nuclear
matter: $\rhoB=0.148$ fm$^{-3}$ (confer, table~2.).
\end{table}

All the hadron masses and nonlinear coefficients are related to
properties of symmetric nuclear matter, such as binding energy, $K$ and
$a_4$, because the chiral breaking mechanism determines nonlinear
interactions in terms of hadron masses and coupling constants, $g$ and
$g_{\omega}$.  Consequently, the mass of $\sigma$-meson, $m_{\sigma}$,
is related to the binding energy of symmetric nuclear matter (${\cal
E}/\rhoB - M = -15.75$ MeV, at $\kfermi = 1.30$ fm$^{-1}$) and adjusted
self-consistently.  The incompressibility is calculated by
\begin{equation}
K = 9\rhoB \frac{\partial^2 {\cal E}}{\partial \rhoB^2} = 9\rhoB \biggl( \frac{\partial \mu}{\partial \rhoB} \biggr) \label{eqn:K} \ ,
\end{equation}
where $\mu$ is the chemical potential and equal to the Fermi energy,
$\mu = E(\kfermi)$, because the current chiral mean-field approximation
is thermodynamically consistent and Landau's hypothesis for
quasiparticles is maintained exactly.  The symmetry energy is calculated
by
\begin{equation}
a_4 = \frac{1}{2}\rhoB \biggl[ \biggl[ \frac{\partial^2 {\cal E}}{\partial \rho_3^2} \biggr]_{\rhoB} \biggr]_{\rho_3 = 0} \label{eqn:RA4} \ ,
\end{equation}
where $\rho_3$ is the difference between the proton and neutron density:
$\rho_3 = \rho_p - \rho_n = (k^3_{F_p} - k^3_{F_n})/3\pi^2$ at a fixed
baryon density, $\rhoB = \rho_p + \rho_n = 2 \kfermi^3/3\pi^2 $.

The coupling constants and effective masses of hadrons, Fermi-liquid
properties of symmetric nuclear matter are listed in the table~1.  The
effective masses of mesons are shown in Fig.~4: $M^{\ast}_N/M_N \sim
0.60$, $m_{\sigma}^{\ast}/m_{\sigma} \sim 1.09$,
$m_{\omega}^{\ast}/m_{\omega} \sim 1.04$, at saturation density.  The
effective mass of nucleon, $M^{\ast}_N/M_N \sim 0.60$, would be
considered to produce a hard EOS and large masses of neutron stars in
nonchiral mean-field approximations, but the chiral mean-field
approximation produces a softer EOS.

The incompressibility and symmetry energy are shown in Fig.~5 and
Fig.~6, respectively, and they are $K = 371$ MeV and $a_4 = 17.4$ MeV,
at saturation density.  These observables are expected to be, $K \sim
300$ MeV and $a_4 \sim 30$ MeV, in the nonchiral, nonlinear
($\sigma,\omega,\rho$) mean-field approximation~$\cite{SCH}$.  One can
notice that $\rho$-meson contribution would be important when $a_4$ in
the nonchiral $(\sigma,\omega,\rho)$ is compared to that of chiral
($\sigma,\omega$) in Fig.~6.  Hence, in order to examine calculations
quantitatively, the chiral ($\sigma,\pi,\omega$) model must be extended
to the chiral ($\sigma,\pi,\omega,\rho$) model~$\cite{S92}$, which is
expected to clarify chiral hadronic models.
\begin{figure}[htpb]
\begin{minipage}[t]{0.48\textwidth}
\begin{center}
\includegraphics[height=.25\textheight]{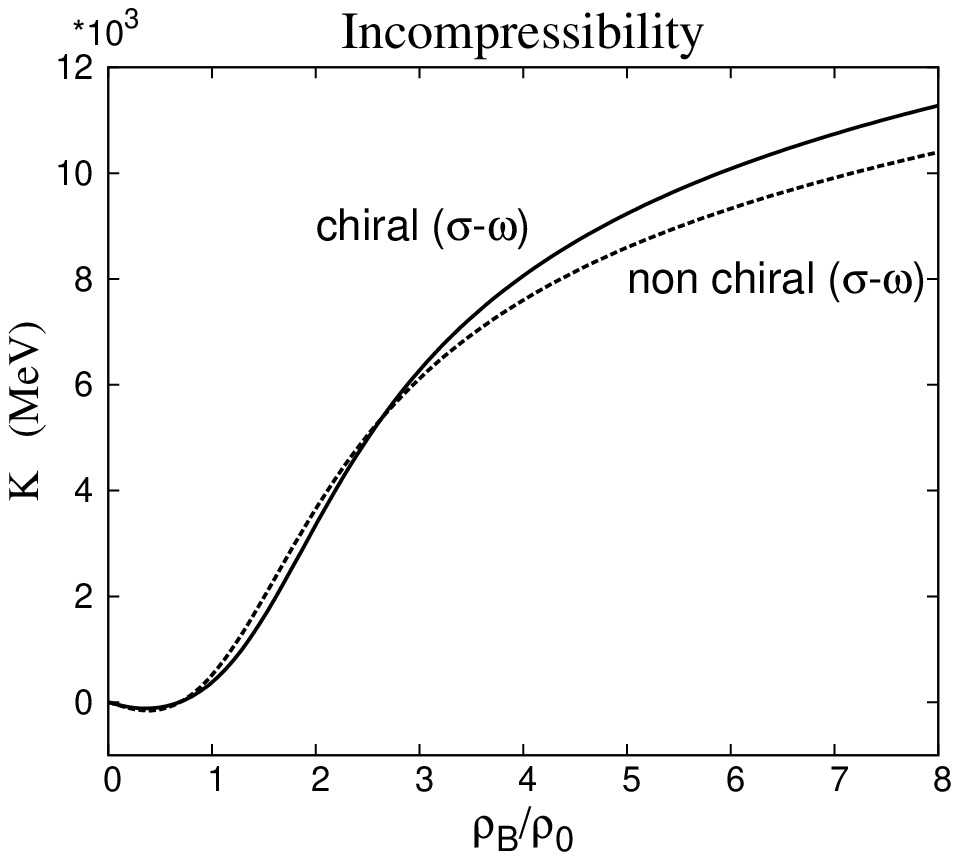}
\label{fig5}
\end{center}\vspace{-0.5cm}
Fig.5.  Incompressibilities in the nonchiral and chiral ($\sigma,\omega$) mean-field approximations.  The effect of
chiral symmetry to incompressibilities is not significant around saturation but important at high densities.
approximation.
\end{minipage}
\hspace{0.5cm}
\begin{minipage}[t]{0.48\textwidth}
\begin{center}
\includegraphics[height=.25\textheight]{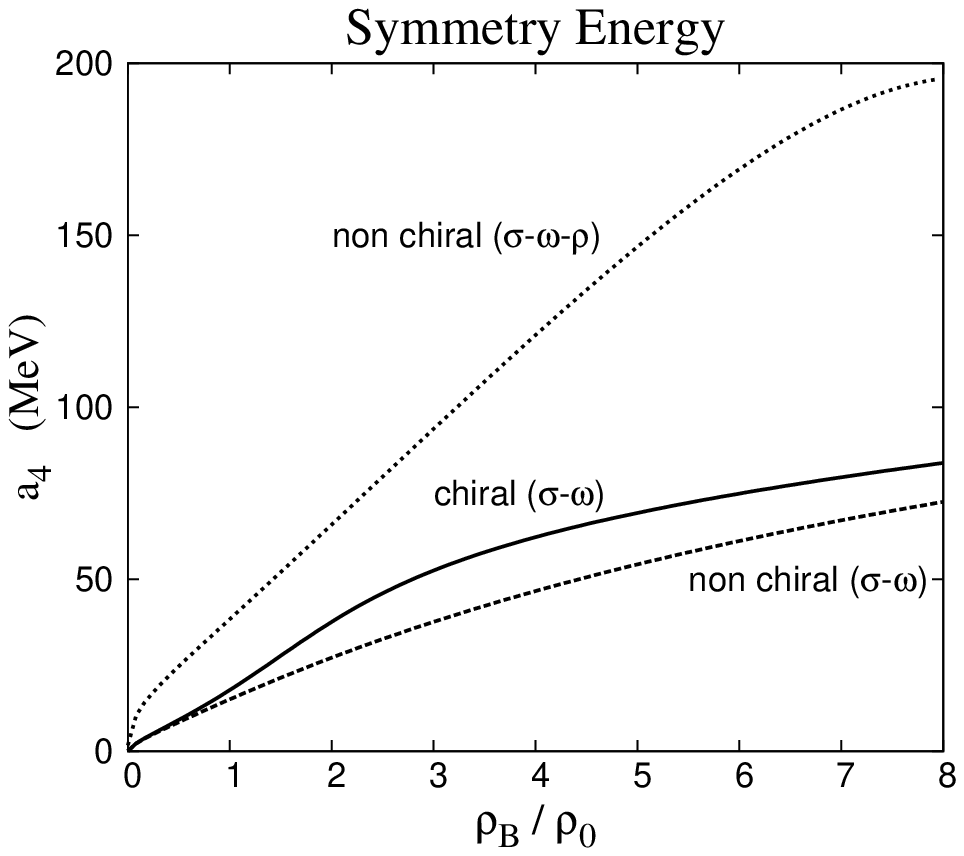}
\label{fig6}
\end{center}\vspace{-0.5cm}
Fig.6.  Symmetry energies in the nonchiral ($\sigma,\omega,\rho$) isospin asymmetric matter, and nonchiral,
chiral ($\sigma,\omega$) isospin symmetric matter.  The $\rho$-meson contribution is more important for $a_4$.
\end{minipage}
\end{figure}

The mass of $\sigma$ meson is important because all the other
observables, EOS, and masses of neutron stars, depend only on the three
adjustble parameters: $m_{\sigma}$ and coupling constants, $g$ and
$g_{\omega}$.  Therefore, the binding energy of symmetric nuclear matter
(${\cal E}/\rhoB - M = -15.75$ MeV, at $\kfermi = 1.30$ fm$^{-1}$) and
the maximum mass of neutron stars ($M_{\rm max} \simeq 2.50$
M$_{\odot}$) determine the mass of $\sigma$ meson to be $m_{\sigma}
\approx 120.0$ MeV.  In (Hartree) mean-field approximations,
contributions of $\pi$-meson vanish in infinite matter due to
spin-saturation, and hence, $\sigma$-meson compensates for $\pi$-meson
contributions in order to produce the saturation mechanism of symmetric
nuclear matter.  The $\sigma$-meson produces attractive interactions at
low densities with the mass: $m_{\sigma} \approx 120.0$ MeV which is
close to the pion mass.  Moreover, $m_{\sigma} \lesssim m_{\pi}$ is
required to obtain solutions consistent with those of the conserving
nonchiral mean-field approximations.  If one assumes $m_{\sigma} >
m_{\pi}$, solutions are restricted to low densities, but the chiral
mean-field approximation is not appropriate in this case because the
interaction potential $V$ shown in Fig.~1 and~2 becomes unbound and
decreases at high densities.

\section{The vacuum fluctuation corrections and Neutron star properties}

The full relativistic chiral Hartree approximation including vacuum
fluctuation corrections (VFC) is derived in this section and applied to
properties of neutron stars.  The divergent integrals coming from
occupied negative energy Dirac vacuum will be rendered finite by
including appropriate counterterms in the current chiral lagrangian.  By
applying the method discussed in the linear $\sigma$-$\omega$ mean-field
approximation~${\cite{WAL}}$ to the nonlinear $\sigma$-$\omega$-$\rho$
mean-field approximation~${\cite{UEC2}}$, the baryon and meson
propagators, self-energies are defined, and appropriate counterterms
that render divergent integrals finite are introduced.

The baryon propagator in mean-field (Hartree) approximation is supposed
to be~${\cite{WAL}}$:
\begin{equation}
\begin{split}
G_{B}^{H}(k)&=(\gamma _{\alpha }k^{\alpha }+M_{B}^{\ast })\left\{ \frac{1}{k^{2}-M_{B}^{\ast }+i\epsilon }
+\frac{i\pi }{E_{B}^{\ast }(k)}\delta (k^{0}-E_{B}(k))\theta (k_{F_{B}}-|{\bm {k}}|)\right\}\\
&=G_{B}^{F}(k)+G_{B}^{D}(k) \ , \label{eqn:fself}
\end{split}
\end{equation}
where $G_{B}^{F}(k)$, $(B=n,p,\Lambda,\cdots)$, is the propagator for
negative energy Dirac-sea and $G_{B}^{D}(k)$ is for density-dependent
Fermi-sea particles, respectively.  It can be readily shown that energy
density, pressure and self-energies in sec.~2 are computed by assuming
$G^{H}(k) = G^D_B (k)$, and hence, we recalculate $\eqref{eqn:S0}$ by
including $G^{F}(k)$, which requires renormalization of infinities into
physical parameters of the model.  By employing the full propagator
$\eqref{eqn:fself}$ in chiral nonlinear $\sigma$-$\omega$ Hartree
approximation, the vector meson self-energy in eq.~$\eqref{eqn:S0}$
becomes
\begin{equation}
\begin{split}
\Sigma _{\omega }^{\mu }&=-i\frac{g_{\omega }}{m_{\omega
 }^{\ast 2}}\sum_{B}\int \frac{d^{4}k}{(2\pi )^{4}}{\rm Tr}\left[
 i g_{\omega}\gamma _{\mu } G_{B}^{H}(k)\right]\\
&=4i\frac{g_{\omega}}{m_{\omega }^{\ast 2}}\sum _{B}g_{\omega}\int \frac{d^{4}k}{(2\pi )^{4}}\frac{k^{\mu }}{k^{2}-M^{\ast 2}+
i\epsilon }-\frac{g_{\omega}^{2}}{m_{\omega }^{\ast 2}}\rho _{\omega }\delta_{\mu, 0}.\label{eqn:Svec1}
\end{split}
\end{equation}
The first term of vector self-energy $\eqref{eqn:Svec1}$ is a divergent
integral evaluated by using the technique of dimensional regularization
as follows,
\begin{equation}
\Sigma _{\omega }^{\mu }=4i\frac{g_{\omega}}{m_{\omega }^{\ast 2}}\sum_{B}g_{\omega}\int \frac{d^{n}k}{(2\pi )^{4}}
\frac{k^{\mu }}{k^{2}-M_{B}^{\ast 2}
+i\epsilon }-\frac{g_{\omega}^{2}}{m_{\omega }^{\ast 2}}\rho _{\omega }\delta_{\mu, 0} \label{eqn:Svec2},
\end{equation}
where the first term of integration is performed in $n$ dimensions, and
the final result of any calculation will be obtained by taking the
physical limit $n \rightarrow 4$.  The integral $\eqref{eqn:Svec2}$
vanishes by symmetric integration, and the fact indicates that
counterterm corrections (CTC) for the chiral mean-field (Hartree)
approximation are produced only by way of $\phi$ fields.

The counterterms to make the scalar self-energy finite are evaluated by
expanding the full propagator of $G^{H}$ in a power series in the {\it
renormalized} scalar self-energy $\Sigma^{\sigma}$.  Using the Dyson
equation, $G^{H}$ is formally expanded as,
\begin{equation}
\begin{split}
G^{H}(k)&=G^{0}(k)+G^{0}(k)\Sigma ^{s}G^{H}(k)\\
&=\sum ^{\infty }_{m=0}\left[ G^{0}({k})\right]^{m+1}\left[ \Sigma ^{s}\right]^{m} \ ,
\end{split}
\end{equation}
and insertion of this expression into the scalar self-energy produces,
\begin{equation}
\Sigma^{s}_H=i\frac{g_{\sigma}}{m_{\sigma }^{\ast 2}}\sum_B\int \frac{d^{n}q}{(2\pi)^4}{\rm Tr}\left[ \sum _{m=0}^{\infty}\frac{1}{m!}[\Sigma^{s}]^m \frac{\partial^{m}G^{0}(q)}{\partial M^m}\right]-\frac{g^2}
{m_{\sigma}^{\ast 2}}\rho_s^{\ast}+\Sigma_{\rm CTC}^{s}.\label{eqn:selfc}
\end{equation}
It is clearly shown that the terms of $m=0,1,2,3$ in $\eqref{eqn:selfc}$
have divergence when the power counting of $q$ is performed in physical
dimension $n=4$.  These divergences can be removed by including the
counterterm contribution in the lagrangian density:
\begin{equation}
{\cal L}_{\rm CTC}=\alpha_1 \phi +\frac{1}{2!}\alpha_2 \phi^2+\frac{1}{3!}\alpha_3 \phi^3 +
\frac{1}{4!}\alpha_4 \phi^4. \label{eqn:ctc}
\end{equation}
The coefficients of $\alpha_1,\alpha_2,\alpha_3,\alpha_4$ are evaluated explicitly by
dimensional regularization${\cite{WAL}}$.  They are given by
\begin{equation}
\begin{split}
&\alpha_{1}=\frac{g}{4\pi ^{2}}\left\{ \Gamma (1-n/2)+2\ln M_{B}+O(n-4)\right\},\\
&\alpha_{2}=-\frac{g^{2}}{4\pi ^{2}}3M_{B}^{2}\left\{ \Gamma (1-n/2)+2\ln M_{B}+\frac{2}{3}+O(n-4)\right\},\\
&\alpha_{3}=\frac{g^{3}}{4\pi ^{2}}\left\{6M_{B} \Gamma (1-n/2)+12M_{B}\ln M+10M_{B}+O(n-4)\right\},\\
&\alpha_{4}=-\frac{g^{4}}{4\pi ^{2}}\left\{ 6\Gamma (1-n/2)+12\ln M_{B}+22+O(n-4)\right\}. \label{eqn:CT1}
\end{split}
\end{equation}
The lagrangian density, ${\cal L}_{\rm CTC}$, is related to the
self-energy $\Sigma_{\rm CTC}^{s}$ by the functional derivative as,
\begin{equation}
\Sigma_{\rm CTC}^{s}=-\frac{g}{m_{\sigma}^{\ast 2}}\frac{\delta {\cal L}_{\rm CTC}}{\delta \phi_{0}},
\end{equation}
and the full self-energy is finally calculated as,
\begin{equation}
\begin{split}
\Sigma_{H}^{\sigma }=&i\frac{g}{m_{\sigma }^{\ast
 2}}\sum_{B}\frac{g}{2\pi ^{2}}\Big[ M_{B}^{\ast 3}\ln
 \frac{M_{B}^{\ast }}{M_{B}}-M_{B}^{2}(M_{B}^{\ast }-M_{B})\\
&\hspace{2.5cm}-\frac{5}{2}M_{B}(M_{B}^{\ast }-M_{B})^{2}-\frac{11}{6}(M_{B}^{\ast }-M_{B})^{3}\Big]
-\frac{g^{2}}{m_{\sigma }^{\ast 2}}\rho _{s}^{\ast}.
\end{split}
\end{equation}

The full energy density is calculated by energy-momentum tensor and
$\eqref{eqn:ctc}$, $\eqref{eqn:CT1}$ as,
\begin{equation}
{\cal E}_{\rm Dirac }(M_{B}^{\ast })=\frac{2\pi ^{n/2}}{(2\pi )^{4}}\Gamma (-n/2)M_{B}^{\ast n}\ ,
\end{equation}
and the vacuum expectation value of energy density defined in the limit
$\kfermi \rightarrow 0$ is given by,
\begin{equation}
{\cal E}_{\rm Dirac }(M_{B})=\frac{2\pi ^{n/2}}{(2\pi )^{4}}\Gamma (-n/2)M_{B}^{n}.
\end{equation}
The finite vacuum fluctuation correction to energy density is determined
from $\eqref{eqn:ctc}$ as $(-\left<\psi|{\cal L}_{\rm
CTC}|\psi\right>)$, and it is calculated as:
\begin{equation}
\begin{split}
\Delta {\cal E}_{\rm VFC}=&{\cal E}_{\rm Dirac}(M_{B}^{\ast })-{\cal E}_{\rm Dirac}(M_{B})-\alpha _{1}\phi -\frac{1}{2!}\alpha _{2}\phi ^{2}
-\frac{1}{3!}\alpha _{3}\phi ^{3}-\frac{1}{4!}\alpha _{4}\phi ^{4}\\
=&-\frac{1}{8\pi ^{2}}\sum_{B}\Big[ M_{B}^{\ast 4}\ln \left( \frac{M_{B}^{\ast }}{M_{B}}\right) +M_{B}^{3}(M_{B}-M_{B}^{\ast })-\frac{7}{2}M_{B}^{2}(M_{B}-M_{B}^{\ast })^{2}\\
&\hspace{3.5cm}+\frac{13}{3}M_B(M_B-M_B^{\ast})^3-\frac{25}{12}(M_B-M_B^{\ast})^4\Big]
\end{split}
\end{equation}
and pressure is given by $\Delta p_{\rm VFC}=-\Delta {\cal E}_{\rm
VFC}$, which is obtained by energy-momentum tensor as: $p =
\frac{1}{3}\left< T^{ii} \right>$, ($i=x,y,z$).  The VFC gives repulsive
contributions for all densities.  The model parameters, $m_{\sigma}$,
$g$ and $g_{\omega}$ must be adjusted and fixed to reproduce
saturation of nuclear matter, where pressure $p = 0$ and ${\cal E}/\rhoB
= E(\kfermi)$ must be satisfied.
\begin{figure}[htpb]
\begin{center}
\includegraphics[height=.25\textheight]{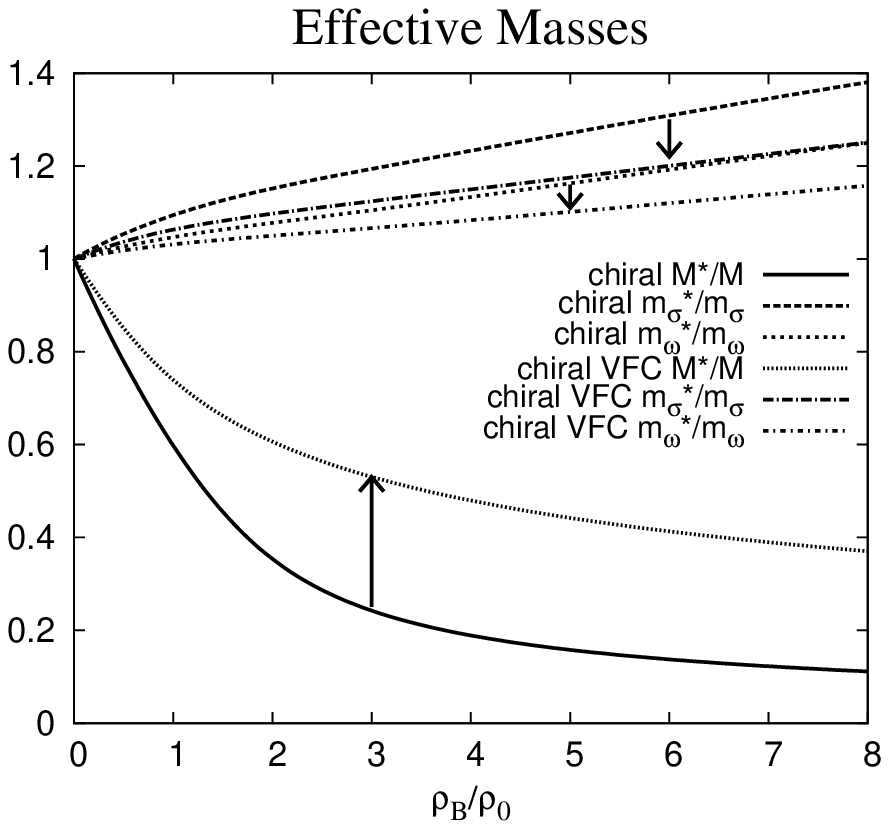}
\label{fig7a}
\end{center}\vspace{-0.5cm}
Fig.7.  The vacuum fluctuation corrections to effective masses in the chiral model.
\end{figure}

The effective masses of baryons and mesons including VFC are shown in
Fig.~7, and at saturation density, they are $M^{\ast}_N/M_N \sim 0.74$,
$m_{\sigma}^{\ast}/m_{\sigma} \sim 1.06$, $m_{\omega}^{\ast}/m_{\omega}
\sim 1.03$; meson effective masses are almost unity around saturation.
The baryon effective mass increases slightly at saturation, which
produces a softer EOS at high densities and decreases the masses of
neutron stars.  The scalar source is decreased a little by VFC, and
accordingly, other fields are similarly decreased by self-consistent
relations required by thermodynamic consistency.  The coupling constants
and effective masses of hadrons, Fermi-liquid properties of symmetric
nuclear matter including VFC are listed in the table~2.

The incompressibility and symmetry energy with VFC are shown in Fig.~8
and Fig.~9.  These Fermi-liquid properties are almost similar at
saturation density, but incompressibility, $K$, is softened at high
densities.  This character shows that the effect of VFC is noticeable at
high densities, but not so important at low densities. The symmetry
energy including VFC gives similar results as discussed in sec.~3, and
one can check from the Fig.~9 that the dominant contribution to $a_4$
should be expected from $\rho$-meson contributions.  The Fock-exchange
corrections produce important contributions to $a_4$ and
$K$~${\cite{UPR}}$.  Since chiral symmetry breaking model confines
coupling constants strictly, it is important to extend chiral models
with $\rho$-meson to the conserving, chiral Hartree-Fock and Brueckner
HF approximations.
\begin{figure}[htpb]
\begin{minipage}[t]{0.48\textwidth}
\begin{center}
\includegraphics[height=.25\textheight]{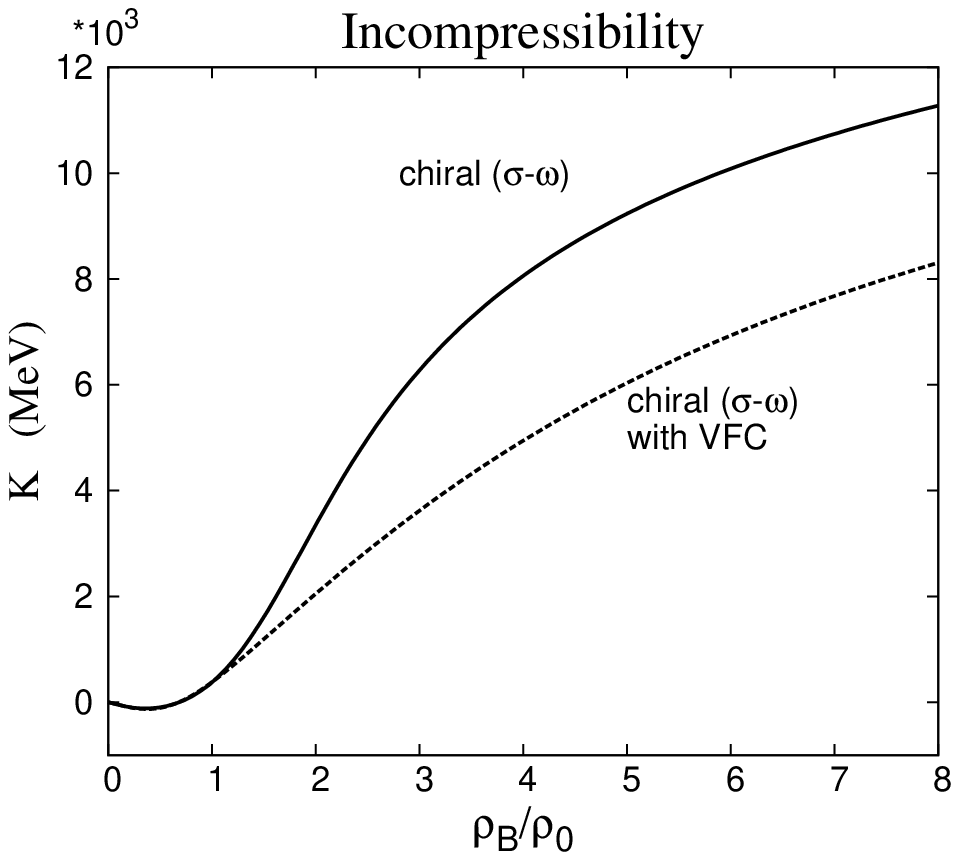}
\label{fig8}
\end{center}\vspace{-0.5cm}
Fig.8.  The vacuum fluctuation corrections to incompressibility in the
chiral model.
\end{minipage}
\hspace{0.5cm}
\begin{minipage}[t]{0.48\textwidth}
\begin{center}
\includegraphics[height=.25\textheight]{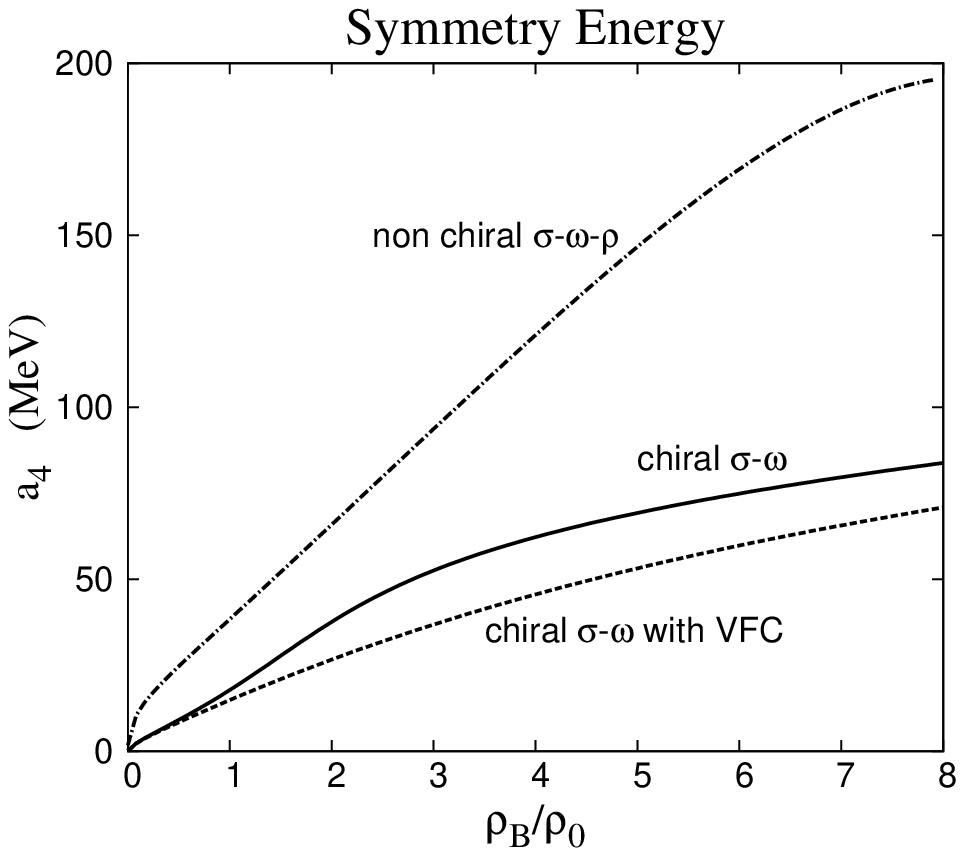}
\label{fig9}
\end{center}\vspace{-0.5cm}
Fig.9.  The vacuum fluctuation corrections to symmetry energy.  The
 nonchiral ($\sigma,\omega,\rho$) calculation is listed for
commparison.
\end{minipage}
\end{figure}
\begin{figure}[b]
\begin{center}
\includegraphics[height=.25\textheight]{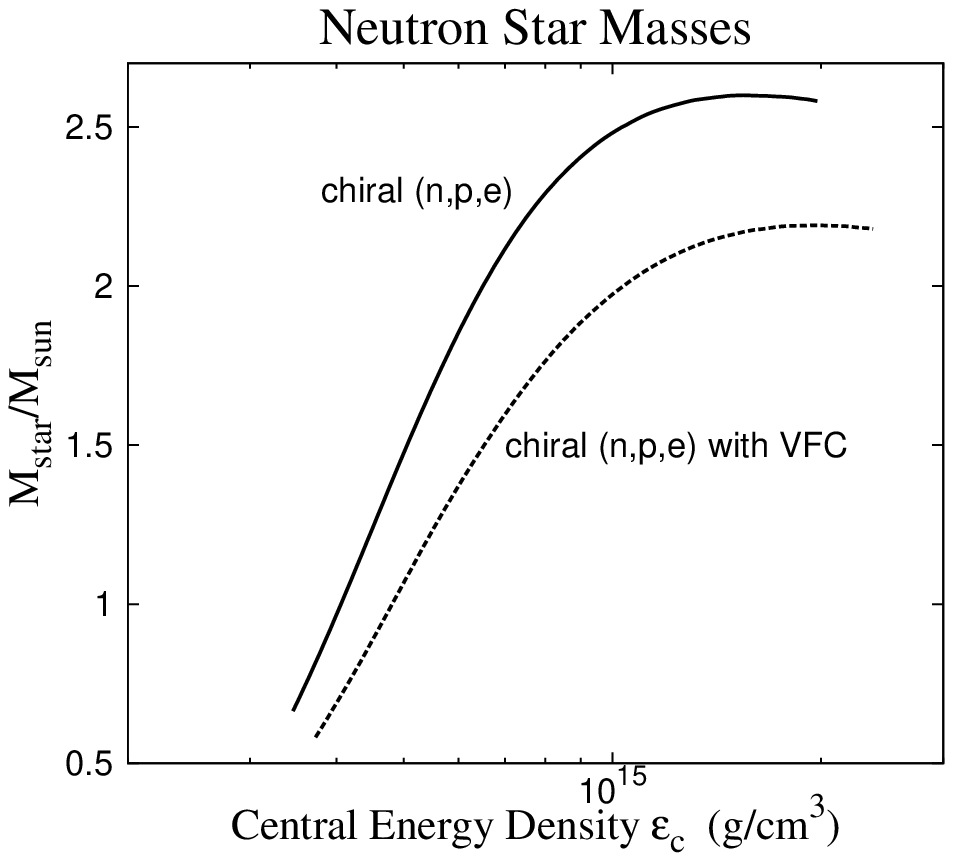}
\label{fig10}
\end{center}\vspace{-0.5cm}
Fig.10.  The masses of neutron stars in the chiral mean-field
approximation, with or without VFC.
\end{figure}
The phase transition from $\beta$-equilibrium ($n,p,e$) to
($n,p,\Lambda,e$) or ($n,p,\Sigma^{-},e$) matter is discussed in the
article~$\cite{SCH}$.  The hyperon-onset densities depend explicitly on
nucleon-hyperon coupling ratios, $r^{\omega}_{HN} = g_{\omega
H}/g_{\omega N}$ and $r^{\sigma}_{HN} = g_{\sigma H}/g_{\omega N}$
$(H=\Lambda$ or $\Sigma^{-})$, and they are given by
\begin{equation}
r^{\omega}_{H N} = \frac{m_{\omega}^{\ast 2}}{g_{\omega N}^{\ } g_{\omega N}^{\ast} \rho_{\omega}}
\Bigl( \frac{g_{\sigma H}}{g_{\sigma N}^{\ast}} (M_N - M_N^{\ast} ) + \alpha_H \Bigr)
= \frac{m_{\omega}^{\ast 2}}{g_{\omega N}^{\ } g_{\omega N}^{\ast} \rho_{\omega}}
\Bigl( M_H - M_H^{\ast} + \alpha_H \Bigr) \ , \label{eqn:ro}
\end{equation}
where $\rho_{\omega} = \rho_p + \rho_n$, and $g_{\omega N}^{\ast}$ is a
density-dependent coupling constant; $\alpha_H$ is the lowest binding
energy of a hyperon.  The coupling ratios are required to be
$r^{\omega}_{\Lambda N} \sim 1.0$ and $r^{\omega}_{\Sigma^{-} N} \sim
1.0$ in the nonchiral, nonlinear ($\sigma,\omega,\rho$) mean-field
approximation in order to obtain optimum empirical values of symmetric
nuclear matter and neutron stars.  If the chiral symmetry breaking is
applied to phase transitions from ($n,p,e$) to ($n,p,\Lambda,e$) or
($n,p,\Sigma^{-},e$) matter, it supports the results that coupling
ratios should be $r^{\omega}_{HN} \sim 1.0$, which is explained as
follows.  In ($\sigma,\pi,\omega$) chiral symmetry breaking models,
$\sigma$-meson generates the mass of nucleon in the new ground state:
$\sigma \rightarrow \sigma_0 + \phi$ and $\sigma_0 = -M_N/g_{\sigma N}$.
Let us include baryons ($n,p,\Lambda,\Sigma^{-},\cdots$) into the
lagrangian~$\eqref{eqn:lag}$ and $\sigma$-hyperon coupling constants are
$g_{\sigma n},g_{\sigma p},g_{\sigma \Lambda}, g_{\sigma
\Sigma{-}},\cdots$, respectively.  Suppose that the ground state
expectation value $\sigma_0$ is equipartitioned to baryons in the new
ground state after chiral symmetry breaking.  Then, one obtains
$-M_n/g_n = -M_p/g_p = -M_{\Lambda}/g_{\Lambda} =
-M_{\Sigma^{-}}/g_{\Sigma^{-}}$, and it results in,
\begin{equation}
r^{\sigma}_{pn} = \frac{g_{\sigma p}}{g_{\sigma n}} = \frac{M_p}{M_n}, \quad
r^{\sigma}_{\Lambda n} = \frac{g_{\sigma \Lambda}}{g_{\sigma n}} = \frac{M_{\Lambda}}{M_n}, \quad
r^{\sigma}_{\Sigma^{-} n} = \frac{g_{\sigma \Sigma^{-}}}{g_{\sigma n}} = \frac{M_{\Sigma^{-}}}{M_n}.
\end{equation}
These values are consistent with those considered appropriate in the
nonlinear ($\sigma,\omega,\rho$) conserving mean-field approximation.

The masses of neutron stars are calculated by using
Tollman-Oppenfeimer-Volkoff (TOV) equation, energy density and pressure
obtained in sec.~3 and sec.~4.  They are shown in Fig.~10 as a function
of a central energy density, ${\cal E}_{c}$.  The vacuum fluctuation
correction softens EOS and reduces the maximum mass of neutron stars
about 20 \%.  It should be noticed that the conserving nonlinear,
nonchiral $\sigma,\omega$ mean-field approximation~$\cite{UEC2}$
reproduces similar results for $K$, $a_4$, $M_{\rm star}$, and
$g_{\sigma}$, $g_{\omega}$ and nonlinear coefficients, when $m_{\sigma}
= 120.0$ MeV is assumed.  Hence, the chiral symmetry breaking mechanism
provides a consistent method to understand solutions to nonchiral,
nonlinear mean-field models. The result indicates that $\rho$-meson is
necessary to obtain reasonable results for properties of Fermi-liquid
and neutron stars.  The analysis with chiral ($\sigma,\pi,\omega,\rho$)
model~$\cite{S92}$ is needed to extract quantitative results.
\begin{table}[]
\noindent{Table~2.\quad Coupling constants and Fermi-liquid properties
 of nuclear matter with VFC}
\begin{center}
\arrayrulewidth=0.5pt
\doublerulesep=0pt
\begin{tabular}{ccccc} \cline{1-3}
$g$   & $g_{\omega}$  & $m_{\sigma}$  \\
1.972 & 10.2235       & 120.0      &  \\ \cline{1-3}
\hline \hline
$\Mstar_N/M_N$ & $m_{\sigma}^{\ast}/m_{\sigma}$ & $m_{\omega}^{\ast}/m_{\omega}$ & $K$ (MeV) & $a_4$ (MeV) \\
 0.74 & 1.06 & 1.03 & 383 & 14.8 \\
\hline
$M_{max}$ & ${\cal E}_{c}$ & $I$ & $R$ (km) \\
 2.19 & 1.88 & 249 & 11.6 \\
\cline{1-4}
\end{tabular}
\end{center}
The result indicates that $\rho$-meson is necessary to obtain reasonable
results for properties of Fermi-liquid and neutron stars.  The analysis
with chiral ($\sigma,\pi,\omega,\rho$) model~$\cite{S92}$ is needed to
extract quantitative results.  $M_{max}$ is the maximum mass in the
solar mass unit $({\rm M}_{\odot})$ and ${\cal E}_{c}(10^{15}{\rm
g/cm^{3}})$ is the central energy density; $I$ is the inertial mass
$({\rm M}_{\odot}{\rm km^{2}})$ and $R ({\rm km})$ is the radius of a
$(n,p,e)$ asymmetric neutron star.
\end{table}

\section{Concluding remarks}

In the current extended chiral mean-field model, all the masses of
baryons and mesons are produced through chiral symmetry breaking of
nonlinear interaction potential, and adjustable free parameters are
limited to $m_{\sigma}$, $g$ and $g_{\omega}$, after hadron masses are
identified and fixed in the nuclear domain, {\it e.g.} $M_N = 939.0$,
$m_{\pi} = 139.0$ and $m_{\omega} = 783.0$ MeV.  The constraints to the
chiral mean-field approximation are properties of saturation (${\cal
E}/\rhoB - M = -15.75$ MeV, at $\kfermi = 1.30$ fm$^{-1}$) and the
maximum mass of isospin-asymmetric neutron stars ($M_{\rm max}(n,p,e)
\lesssim 2.50$ M$_{\odot}$).  The mass of $\sigma$-meson is determined
to maintain the constraints and given by $m_{\sigma} \simeq 120$ MeV,
which is also necessary so that the interaction potential $V$ is
positive and bounded at high densities.  The chiral mean-field
approximation indicates that a scalar particle close to the mass of
$\pi$-meson should be needed to produce saturation of nuclear matter.

The effective masses of nucleon and mesons, $M_N^{\ast},
m_{\sigma}^{\ast}, m_{\omega}^{\ast}$, are similar to those derived from
nonchiral, nonlinear ($\sigma,\omega$) mean-field approximation.  The
effective mass of nucleon $M_N^{\ast}/M_N$ monotonically decreases, but
effective masses of mesons are, $1.0 \lesssim m_{\sigma}^{\ast}/m_{\sigma},
m_{\omega}^{\ast}/m_{\omega}$, at or around saturation density.  The vacuum
fluctuation corrections exhibit repulsive effects for all densities, but
after adjusting coupling constants to reproduce properties of saturation
and neutron stars, it is examined that the effect of VFC is less
significant at saturation than that at high densities.  The effect of
nonlinear interactions is more important than that of VFC in the Hartree
approximation.  The similar conclusion is also obtained in the
nonchiral, nonlinear ($\sigma,\omega,\rho$) mean-field approximation.
As shown in Fig.~6, $\rho$-meson gives noticeable contributions, and so,
the chiral nonlinear $(\sigma,\bm{\pi},\omega)$ mean-field approximation
should be extended and examined by including $\rho$-meson.

The nonchiral, nonlinear $(\sigma,\omega,\rho)$ mean-field
approximations have many adjustable nonlinear coupling constants.  The
nonlinear coupling constants have upper bound restricted by
self-consistent conditions to approximations and properties of
saturation and neutron stars~$\cite{UEC2}$, which is expected as a
manifestation of {\it naturalness} of nonlinear
coefficients~$\cite{RBH}$.  The current chiral mean-field approximation
determines all the nonlinear constants in terms of three adjustble
parameters: $m_{\sigma}$, $g$ and $g_{\omega}$.  The masses of mesons,
$m_{\pi}$ and $m_{\omega}$, are identified and fixed by experimental
values, after the chiral symmetry breaking.  The nonlinear constants
expressed by $m_{\sigma}$, $g$ and $g_{\omega}$ support the properties
of naturalness and the bounded values of nonlinear constants given by
nonchiral, nonlinear ($\sigma,\omega,\rho$) mean-field approximation.
The self-consistent and optimum solutions to the nonchiral, nonlinear
($\sigma,\omega$) mean-field approximation with $m_{\sigma} = 120.0$ MeV
become similar to those of the chiral ($\sigma,\omega$) mean-field
approximation, and so, it suggests that chiral symmetry serves to
restrict solutions to nonlinear mean-field approximations.

Because the chiral-symmetry breaking relates nonlinear coefficients with
hadron masses, the chiral mean-field approximation suggests that
nucleon-proton, nucleon-hyperon coupling ratios be given by ratios of
hadron masses, such that $r^{\sigma}_{\Lambda N} = M_{\Lambda}/M_N
\approx r^{\omega}_{\Lambda N}$ and $r^{\sigma}_{\Sigma N} =
M_{\Sigma}/M_N \approx r^{\omega}_{\Sigma N}$.  It is remarkable that
the values of coupling ratios are consistent with those obtained by the
condition at hyperon-onset density, which is determined by the
requirement of thermodynamic consistency at saturation of hyperon
matter~$\cite{SCH}$.  The coupling ratios produce reasonable
density-dependent properties of nuclear matter and neutron stars in the
calculation of conserving nonchiral, nonlinear ($\sigma,\omega,\rho$)
mean-field approximation.  On the contrary, the coupling ratios given by
the SU(6) quark model for vector coupling constants~$\cite{JSI,GSY}$ are
expected to be $r^{\omega}_{\Lambda N} = 2/3$ and $r^{\omega}_{\Sigma N}
= 2/3$, but the ratios do not generate consistent results for properties
of nuclear and neutron matter.  The chiral symmetry breaking is useful
to understand relations among nonlinear coupling constants for nonlinear
mean-field models.

The effect of VFC is not prominent at saturation density compared with
that of nonlinear interactions, but VFC softens EOS at high densities
and the EOS would be softened further when hyperons are generated; the
fact is also consistent with the result derived from the nonchiral,
nonlinear ($\sigma,\omega,\rho$) mean-field approximation~$\cite{STU}$.
Since chiral symmetry breaking clarifies relations among nonlinear
interactions, it is important to understand how hyperon-onset densities,
binding energy and saturation properties of hyperon matter, masses of
hadron and hadron-quark stars would be modified by chiral models of
hadrons.  The chiral symmetry breaking mechanism helps us understand
physical meanings of chiral symmetry to masses and coupling constants of
hadrons; quantitative analysis in terms of the chiral symmetry breaking
may help us understand if $\sigma$-meson is a real particle state or a
virtual state characteristic to Hartree (mean-field) approximation.  The
problems of high-energy hadron scatterings and properties of infinite
matter such as hadron-quark stars suggest that hadronization from QCD,
phase transition from bound state hadrons to quark matter can be one of
important topics in the near future.  However, the quantitative analysis
in terms of both quantumhadrodynamics (QHD) and QCD is necessary.  The
current self-consistent chiral mean-field model and other chiral models
should be extended to more sophisticated approximations, such as
conserving HF and BHF approximations in order to obtain consistent
results.


{\bf Acknowledgement}

We would like to acknowledge the Research Center for Nuclear Physics
(RCNP) at Osaka University; especially, Dr. M.~Valverde and Dr. J.~Hu
for their constructive opinions and supports.


\end{document}